\documentclass[aps,prl,twocolumn,a4paper]{revtex4}
\pdfoutput=1

\newcommand{\kB}{k_{\textrm{B}}}

\usepackage{geometry}
\usepackage{color}
\usepackage[latin1]{inputenc}
\usepackage{graphicx}
\usepackage{amsmath,amssymb}

\usepackage{times}

\usepackage{textcomp}
\usepackage{amsfonts}
\usepackage{graphicx}
\usepackage{color}
\usepackage[latin1]{inputenc}
\usepackage{braket}
\usepackage{units}
\usepackage[svgnames]{xcolor}
\usepackage[
pdftitle={Emergence of a Turbulent Cascade in a Quantum Gas},    
pdfauthor={Small Machine},     
pdfsubject={Turbulence},   
pdfkeywords={Turbulence} {Turbulence cascade} {Bose-Einstein condensate}, 
colorlinks=True,linkcolor=DarkSlateBlue,citecolor=DarkBlue,urlcolor=DarkBlue,
	pdfstartview=FitH,bookmarks=False,pdfpagemode=UseNone
]{hyperref}

\begin{document}

\title{Emergence of a Turbulent Cascade in a Quantum Gas}

\date{\today}

\author{Nir Navon, Alexander L. Gaunt, Robert P. Smith and Zoran Hadzibabic}

\affiliation{Cavendish Laboratory, University of Cambridge, 
J. J. Thomson Avenue, Cambridge CB3 0HE, United Kingdom}

\maketitle

{\bf 
In the modern understanding of turbulence, a central concept is the existence of cascades of  
excitations from large to small lengthscales, or vice-versa. 
This concept was introduced
in 1941 by Kolmogorov and Obukhov~\cite{kolmogorov1941local,obukhov1941distribution}, and the phenomenon has since been observed in a variety of systems, including interplanetary plasmas~\cite{sorriso2007observation}, supernovae~\cite{mosta2015large}, ocean waves~\cite{hwang2000airborne}, and financial markets~\cite{ghashghaie1996turbulent}. 
Despite a lot of progress, quantitative understanding of turbulence remains a challenge due to the interplay of many lengthscales that usually thwarts theoretical simulations of realistic experimental conditions. Here we observe the emergence of a turbulent cascade in a weakly interacting homogeneous Bose gas, a quantum fluid that is amenable to a theoretical description on all relevant lengthscales. We prepare a Bose--Einstein condensate (BEC) in an optical box~\cite{gaunt2013bose}, drive it out of equilibrium with an oscillating force that pumps energy into the system at the largest lengthscale, study the BEC's 
nonlinear response to the periodic drive, and observe a gradual development of a cascade characterised by an isotropic  power-law distribution in  momentum space. 
We numerically model our experiments using the Gross--Pitaevskii equation (GPE) and find excellent agreement with the measurements. Our experiments establish the uniform Bose gas as a promising new platform for investigating many aspects of turbulence, including the interplay of vortex and wave turbulence and the relative importance of quantum and classical effects. 
}

Compared to classical fluids, superfluids present fascinating peculiarities such as irrotational and frictionless flow, which raises fundamental questions about the character of turbulent cascades~\cite{paoletti2011quantum,chesler2013holographic}. Numerous experiments on quantum-fluid turbulence have been performed with liquid helium, exploring both vortex \cite{maurer1998local,walmsley2008quantum,bradley2011direct,paoletti2011quantum} and wave turbulence~\cite{Ganshin:2008,Abdurakhimov:2012,Kolmakov:2014},  
but their theoretical understanding is hampered by the strong interactions that make first-principle descriptions intractable. 
The situation is {\it a priori} simpler for an ultracold weakly interacting Bose gas, which is often accurately described by the GPE for the complex-valued matter field $\psi({\bf r},t)$~\cite{kagan1997evolution}.
This equation is widely used to model turbulence in quantum fluids~\cite{nore1997kolmogorov,berloff2002scenario,kobayashi2005kolmogorov,proment2009quantum,reeves2013inverse}, but the numerical results have been lacking experimental validation.
Experimentally, qualitative evidence for turbulence has been seen in quantum gases~\cite{henn2009emergence,neely2013characteristics,kwon2014relaxation,Tsatsos:2016}, but quantitative comparisons with theory were hindered by the inhomogeneous density resulting from harmonic trapping.  Here we eliminate this problem by studying turbulence in a homogeneous quantum gas.

\begin{figure} [bp]
\includegraphics[width=\columnwidth]{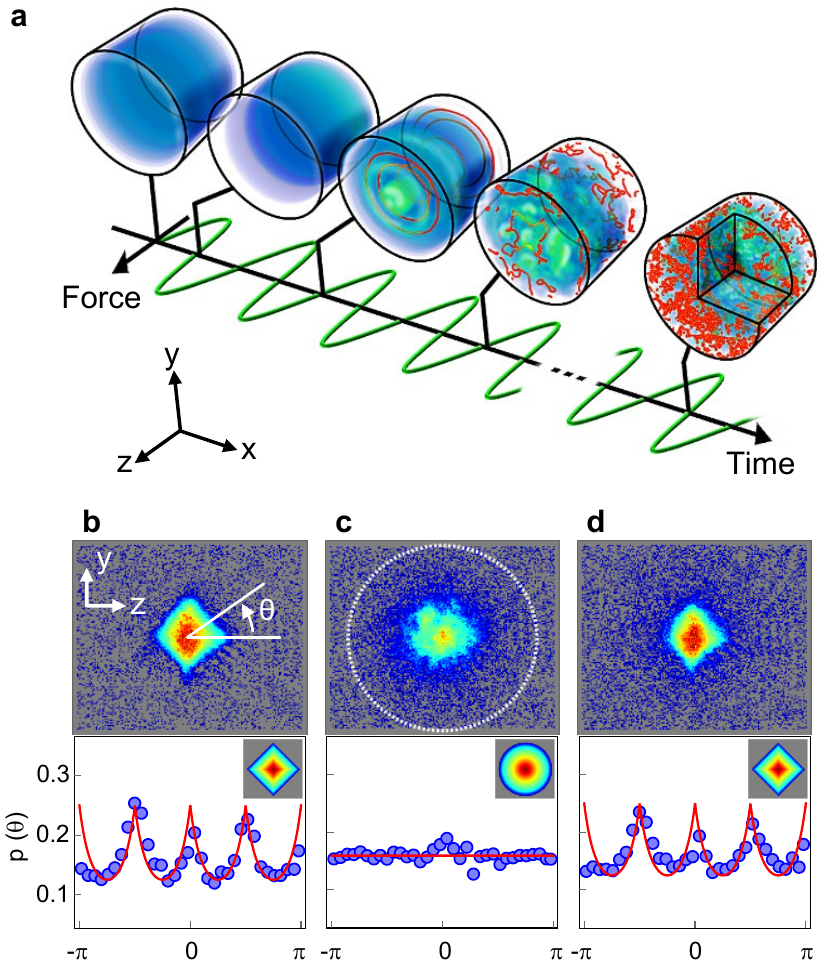}
\caption{
{\bf From unidirectional sloshing to isotropic turbulence.} {\bf a}, Gross-Pitaevskii simulations of a shaken box-trapped Bose gas. Red lines indicate vortices. {\bf b-d}, Experimental absorption images taken along $x$ after 100~ms of time-of-flight expansion, with $N\approx 8\times 10^4$ atoms (upper panels), and corresponding angular distributions $p(\theta)$, averaged over 20 images taken under identical conditions (lower panels). {\bf b}, Initial BEC, {\bf c}, after shaking for $2\,$s at $8\,$Hz with amplitude $\Delta U/\mu \approx 1.2$, and {\bf d}, after the turbulent cloud was allowed to relax for 1.5~s. The dashed circle in {\bf c} corresponds to expansion energy $\kB T_c/2$. In the lower panels, the red lines correspond to the diamond-like and isotropic distributions depicted in the insets.
}
\label{Fig1}
\end{figure} 

\begin{figure*}[t!]
\includegraphics[width=2\columnwidth]{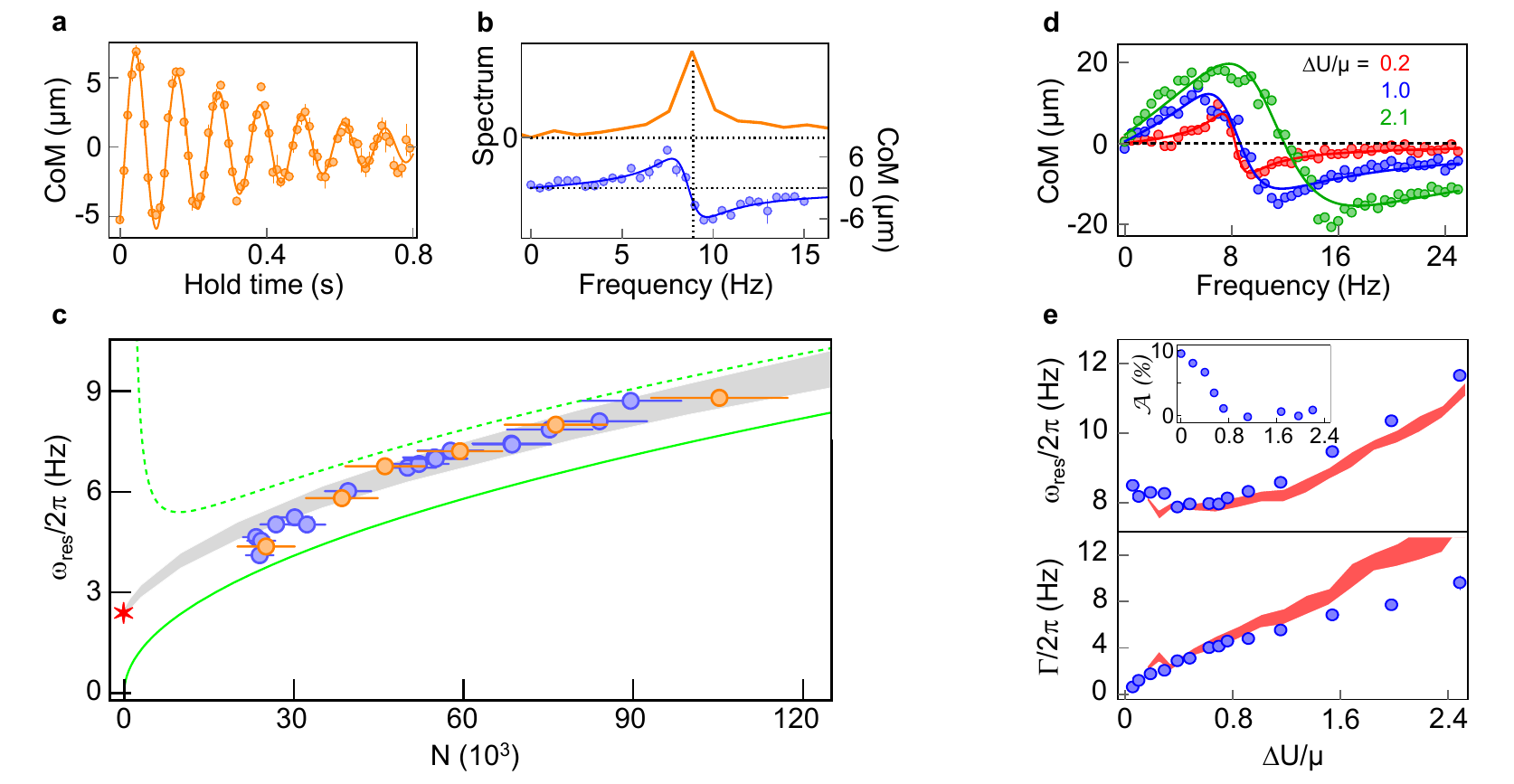}
\caption{
{\bf Route to turbulence: Nonlinear spectroscopy of the BEC's lowest axial mode.} {\bf a-b}, Small-amplitude CoM oscillations for $N=10(1)\times 10^4$. {\bf a}, Free oscillation after a 20-ms kick with $\Delta U/k_B=1.7$~nK. 
{\bf b}, Fourier spectrum of the free oscillation in {\bf a} (orange), and displacement after a whole number of driven oscillations, with driving amplitude $\Delta U/k_B= 0.3$~nK and $t_S=2\,$s (blue). 
{\bf c}, Small-amplitude $\omega_{\rm res}$ versus $N$ for free (orange) and driven (blue) oscillations with the same $\Delta U$ and $t_S$ values as in {\bf a}-{\bf b}. 
Horizontal error bars represent 1-$\sigma$ errors, while the $\omega_{\rm res}$ errors are smaller than the point size. The grey shaded area shows numerical solutions of Bogoliubov equations and the red star is the analytical non-interacting limit. Green lines are based on hydrodynamic approximations (see text). {\bf d-e}, Nonlinear response, for $N=8(1)\times 10^4$. {\bf d}, Driven-oscillation signals as in {\bf b}, for various $\Delta U$. {\bf e}, $\omega_{\rm res}$ and linewidth $\Gamma$ versus $\Delta U$ (blue points), and the corresponding results of GPE simulations (red bands). Inset: ToF anisotropy $\mathcal{A}$ versus $\Delta U/\mu$ for $t_S=4$~s of resonant driving. 
All CoM measurements were done with $t_{\rm ToF}=140$~ms. 
}
\label{Fig2}
\end{figure*}

The basic idea of our experiment is outlined in Fig.~\ref{Fig1}. We prepare a quasi-pure BEC of $^{87}$Rb atoms in a cylindrical optical box~\cite{gaunt2013bose}, and drive it out of equilibrium with a spatially-uniform oscillating force that primarily couples to the lowest, dipole-like axial mode.
Our box has length $L = 27(1) \, \mu$m and radius $R = 16(1)\,\mu$m. For our typical atom number, $N\approx 10^5$, the initial equilibrium BEC has a chemical potential $\mu/\kB \approx 2\,$nK, interaction energy per particle $E_{\rm int}/\kB \approx 1\,$nK, and negligible kinetic energy, while the BEC critical temperature is $T_c \approx 50\,$nK.
The driving force is provided by a magnetic field gradient that creates a potential $U({\bf r})=\Delta U z/L$, where the coordinate $z$ is along the axis of the box (Fig.~\ref{Fig1}a).  
The natural scale for $\Delta U$, separating weak and strong drives, is set by $\mu$.

Numerical simulations in Fig.~\ref{Fig1}a show the microscopic behaviour of a shaken trapped gas, which gradually changes from simple unidirectional sloshing along $z$ to an omnidirectional turbulent flow; in addition to the wave-like motion we observe vortex lines (depicted in red), which are detected by computing the local circulation~\cite{FootnoteSnapshots}. Here the shaking amplitude is $\Delta U/\mu =1$ and the longest shaking time, $t_S = 2\,$s, corresponds to 16 driving periods.

Experimentally, we probe the global properties of the gas by releasing it from the trap and imaging it along a radial direction ($x$) after a long time-of-flight (ToF) expansion, $t_{\rm ToF} \geq 100\,$ms. From the images we extract the cloud's centre of mass (CoM) and momentum distribution. The CoM position reflects the axial in-trap sloshing and the evolution of the momentum  distribution reveals the cascade of excitations from small to large wavevectors $\bf k$, the so-called direct cascade. 

In Fig.~\ref{Fig1}b-d we show a qualitative experimental signature of turbulence with three key examples of ToF images (upper panels) and the corresponding angular distributions of atoms, $p(\theta)$ (lower panels). The initial BEC (Fig.~\ref{Fig1}b) shows an anisotropic expansion, which is driven by the conversion of interaction into kinetic energy, and reflects the shape of the container~\cite{gotlibovych2014observing}. 
In sharp contrast, after sufficiently long shaking the expansion is isotropic (Fig.~\ref{Fig1}c), with $p(\theta)$ showing small fluctuations around $1/(2\pi)$. This is the first qualitative signature of a kinetic-energy dominated turbulent state, in which the long-range coherence of the BEC is destroyed. 
We stress that this highly non-equilibrium state is fundamentally different from an equilibrium non-condensed state, which is also kinetic-energy dominated and displays isotropic expansion. The key point is that in our box trap there is a large separation between the initial $E_{\rm int}$ and $\kB T_c$. This gives us access to the regime where the total (mostly kinetic) energy per particle, $E$, satisfies $E_{\rm int} \ll E \ll \kB T_c$. In this regime, coherence is destroyed in the turbulent state, but the corresponding equilibrium state with the same $E$ is still deeply condensed. 
In Fig.~\ref{Fig1}c, the dashed circle corresponds to expansion energy $\kB T_c/2$, and the average energy of the atoms is clearly much lower; from the second moment of the ToF distribution we get $E \approx 0.12 \, \kB T_c$, which in equilibrium would correspond to a condensed fraction $\eta \approx 0.7$~\cite{schmidutz2014qjt}. Indeed, if we stop shaking and allow the turbulent gas to relax, a BEC reforms (Fig.~\ref{Fig1}d), with the expected $\eta = 0.7(1)$.
 For all our studies of the turbulent state we restrict the shaking amplitude, $\Delta U \lesssim 2\mu$, and time, $t_S \leq 4$~s, so that $E < 0.25 \, \kB T_c$, corresponding to  equilibrium $\eta > 0.5$.

\begin{figure*}[t!]
\includegraphics[width=2\columnwidth]{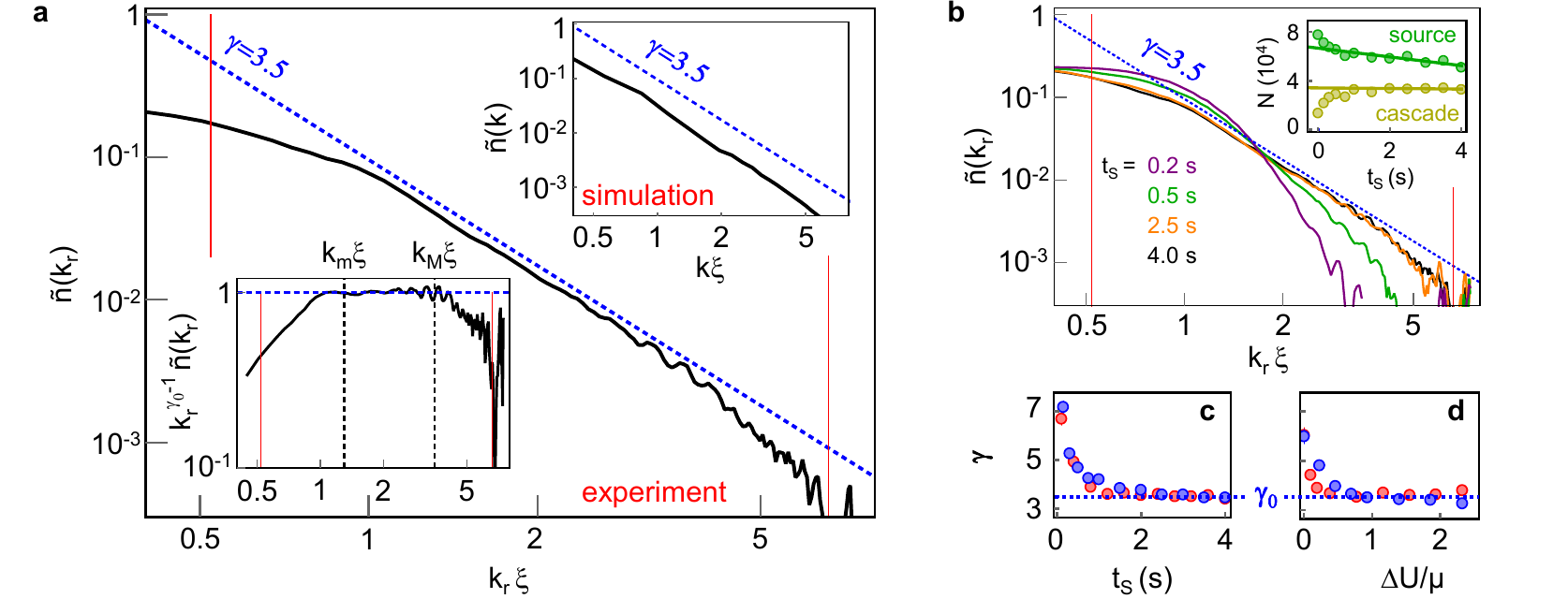}
\caption{
{\bf Development of a turbulent cascade.} {\bf a}, Momentum distribution of the turbulent gas (solid black line), for $N=7(1)\times 10^4$, $\Delta U/\mu = 1.1(1)$, $t_S=4$~s, $\omega/2\pi=9$~Hz and $t_{\rm ToF} = 100\,$ms. 
The vertical red lines indicate the momentum resolution $k_{\rm low}$ (left) and the energy sink at $k_{\rm high}$ (right); the dashed blue line is a guide to the eye, offset from the data for clarity.
Lower inset:  Compensated spectrum $k_r^{\gamma_0 - 1}\tilde{n}(k_r)$ with $\gamma_0=3.5$ (in log-log scale); $k_m$ and $k_M$ define the fitting ranges used in {\bf b}-{\bf d}. Upper inset: Steady-state distribution from GPE simulations, for $\Delta U/\mu=1$. 
{\bf b}, Dynamics of $\tilde{n}(k_r)$ towards the steady state, for $\Delta U/\mu=1.1(1)$.
Inset: Total atom population for $k_r<k_m$ (the low-$k$ `source') in green, and for $k_m<k_r<k_M$ (in the cascade region) in yellow. 
At long times (solid lines) $\dot{N}_{\rm source}=-3.6(1.5)$ atoms/ms, while $\dot{N}_{\rm casc}=-0.2(3)$ atoms/ms is consistent with zero. All populations are corrected for losses due to the collisions with the background gas in the vacuum chamber (see Methods).
{\bf c}, Exponent $\gamma$ versus shaking time in experiment (blue, $\Delta U/\mu = 1.1(1)$) and simulations (red, $\Delta U/\mu = 1$). {\bf d}, Exponent $\gamma$ versus shaking amplitude in experiment (blue) and simulations (red), for $t_S=4\,$s. 
}
\label{Fig3}
\end{figure*} 

To see how pumping energy at the largest lengthscale (with a spatially uniform force) leads to a turbulent cascade, we perform detailed spectroscopy of the lowest-lying axial mode of the BEC (see Fig.~\ref{Fig2}). In contrast to the harmonic trap, where the dipole mode is fixed by the trapping frequency (Kohn's theorem), in the box it depends on interactions, which results in nonlinear behaviour for non-vanishing shaking amplitudes.

We first study the small-amplitude CoM response for various $N$, using both free and driven oscillations. 
In the first method, we pulse on a constant $\Delta U$ for a short time and let the gas oscillate freely in the trap for a variable hold time 
before releasing it and measuring the CoM in ToF (Fig.~\ref{Fig2}a). 
In the Fourier spectrum of this oscillation (orange line in Fig.~\ref{Fig2}b) we see a single strong peak, with no indication that the gradient kick directly couples to other low-lying modes. 
In the second method, we apply a continuous oscillating drive of frequency $\omega$ and amplitude $\Delta U \lesssim 0.5\mu$, and perform ToF measurements after a whole number of drive periods. Similarly to a driven harmonic oscillator, the CoM displacement shows a dispersive line-shape as a function of $\omega$, vanishing on resonance (see Methods). 
 As shown in Fig.~\ref{Fig2}b, the two methods give the same resonant frequency $\omega_{\rm res}$.

In Fig.~\ref{Fig2}c we plot the small-amplitude $\omega_{\rm res}$  versus $N$, and compare the data with various theories.
The hydrodynamic prediction (solid green line) is $\omega_{\textrm{HD}}=\pi c/L$, where $c=\sqrt{\mu/m}$ is the speed of sound and $m$ is the atom mass.
This theory assumes that the healing length, $\xi=\hbar/\sqrt{2m\mu}$, satisfies $\xi \ll L$. It is thus not applicable in the $N\rightarrow 0$ limit, where $\omega_{\rm res} = 3 \pi^2 \hbar/(2mL^2)$ (red star) is given by the splitting of the lowest axial single-particle states. For our largest BEC, $L/\xi \approx 20$, but $\omega_{\rm HD}$ is still observably lower than $\omega_{\rm res}$. 
Interestingly, we empirically find that an upper bound on $\omega_{\rm res}$ (dashed green line) is obtained by simply calculating $\omega_{\rm HD}$ for an effective BEC volume that excludes the region within $\xi$ of the trap walls. Finally, we linearise the GPE around the ground-state BEC solution for our box trap and numerically solve the resulting Bogoliubov equations (see Methods). These solutions are shown as the grey shaded area, which accounts for the experimental uncertainty in the box size. We find excellent agreement with the data, without any adjustable parameters.

In Fig.~\ref{Fig2}d-e we show measurements for driven oscillations with different drive strengths. Increasing $\Delta U$ both shifts and broadens 
the resonance, and both trends are reproduced by our GPE simulations (red bands in Fig.~\ref{Fig2}e); for very large $\Delta U$ the classical-field GPE approximation may gradually break down. The line broadening, seen for any non-zero $\Delta U$, indicates nonlinear coupling to other modes, which provides the route for the transfer of excitations into other directions and a direct cascade.

In the inset of Fig.~\ref{Fig2}e we plot the anisotropy of the ToF expansion, $\mathcal{A}= (1/2) \int \textrm{d}\theta \, \left| p(\theta)- 1/(2\pi) \right|$ (see Methods), for 4~s of resonant driving. For $\Delta U \gtrsim 0.8\,\mu$ we observe the isotropic expansion ($\mathcal{A}\approx 0$) that qualitatively signals turbulence. A key quantitative expectation for an isotropic turbulent cascade is the emergence of a steady-state power-law momentum distribution, $n({\bf k}) \propto k^{-\gamma}$, where $\gamma$ is a constant~\cite{zakharov1992kolmogorov}. Due to the line-of-sight integration in absorption imaging, this corresponds to an in-plane distribution $\tilde{n}(k) \propto k^{-(\gamma -1)}$. 

In Fig.~\ref{Fig3} we present our study of $\tilde{n}(k)$ observed after a resonant drive. An isotropic expansion (from an anisotropic container) necessarily means that the in-trap kinetic energy dominates over the interaction energy, which in turn means that the ToF expansion can provide an accurate measure of the in-trap momentum distribution. Specifically, defining $k_r \equiv m r/(\hbar t_\textrm{ToF})$, where $r$ is distance from the CoM in ToF, $\tilde{n}(k_r)$ should closely correspond to the in-trap $\tilde{n}(k)$ (see Methods, Extended Data Fig.~1). 
Note, however, that this correspondence does not hold for very low momenta, $k_r \lesssim k_\textrm{low} \equiv m L/(\hbar t_\textrm{ToF})$, due to the convolution of the ToF distribution with the initial (in-trap) cloud shape. The highest momentum in our clouds, $k_\textrm{high} \equiv \sqrt{2m U_0}/\hbar$, is set by the trap depth, $U_0 \approx \kB \times 60$~nK, which corresponds to an energy sink.

In Fig.~\ref{Fig3}a we show an example of $\tilde{n}(k_r)$, for $\Delta U/\mu=1.1(1)$ and $t_S=4\,$s 
(black line in the main panel and lower inset), obtained by averaging over 20 images and also performing an azimuthal average. Vertical red lines indicate the $k_\textrm{low}$ and $k_\textrm{high}$ boundaries. 
Away from these boundaries we clearly observe power-law behaviour, with $\gamma\approx 3.5$. This is visually even more striking in the lower inset, where we plot $k_r^{\gamma_0-1} \tilde{n}(k_r)$, with $\gamma_0\equiv 3.5$. 
In the top inset we show the result of GPE simulations (for $\Delta U/\mu=1$), which also exhibits a power-law distribution. Moreover, the experiment and simulations are consistent with the same value of $\gamma$.

In Fig.~\ref{Fig3}b we present the evolution of $\tilde{n}(k_r)$ towards the turbulent steady state, 
as the shaking time is increased.  
In the inset we show (on a linear scale) the total atom populations in the low-$k$ `source' region, $k_r < k_m$ (see Fig.~\ref{Fig3}a), and in the range $k_m < k_r <k_M$,  where the power-law distribution is established in steady state.
Initially there is a net transfer of population from the source to the cascade region. The population growth in the cascade region means that at these early times the population-flux through this $k$-space range is not constant. However, once the steady state is established the population in the cascade $k$-range saturates at a constant value, while the source is still slowly depleted.  This is indeed what is expected for a direct cascade, in which a constant, $k$-independent population-flux passes from the source, through the cascade range, to the high-$k$ sink; note that formally this population flux, for a given energy flux, should tend to zero as the sink is moved towards infinite energy~\cite{zakharov1992kolmogorov}. (For a non-infinite-energy sink, one strictly speaking has a quasi-steady state, since at very long times the source would be too depleted to support a constant-flux cascade.)

We further cross-validate our experiments and first-principle calculations by fitting the cascade exponent $\gamma$ in the range $k_m < k_r < k_M$. 
In Fig.~\ref{Fig3}c we show that, for $\Delta U/\mu \approx 1$, the experiment and simulations exhibit very similar evolution with the shaking time, and reach a steady-state value of $\gamma$ after $t_S \approx 2\,$s. In Fig.~\ref{Fig3}d we plot the measured and simulated $\gamma$ values versus the shaking amplitude for fixed $t_S=4$~s. Here we see that the steady-state value of  $\gamma$ is essentially independent of $\Delta U$, reinforcing the robustness of our conclusions (for small $\Delta U$ the steady state is not reached for $t_S=4$~s; see also inset of Fig.~\ref{Fig2}e).

 
We lastly also discuss our findings in the context of previous theoretical work. 
The $\gamma$ we observe in both experiments and simulations is close to one of the scarce analytical predictions, the Kolmogorov-Zakharov direct-cascade exponent $\gamma=3$, for the weak-wave turbulence of a compressible superfluid~\cite{zakharov1992kolmogorov,KolmogorovSpectrumNote}. 
This prediction is based on an idealised theory that starts with the GPE, but neglects the role of vortices and assumes weak interactions between the waves. 
Our simulations show that vortices are present in the system, but the value of $\gamma$ suggests that they do not play a quantitatively significant role in the turbulent cascade (observed at wavenumbers $k\xi\gtrsim 1$), and the Kolmogorov-Zakharov theory is a reasonable approximation. 
Consistently, in simulations we find that in the relevant $k$-range the compressible-flow contribution to the energy dominates over the incompressible-flow one (see Methods, Extended Data Fig.~2).
The small difference between $\gamma =3.5$ and the approximate $\gamma=3$ could arise due to a number of (interlinked) factors, such as a residual role of vortices, the non-negligible incompressible-flow energy, the fact that in reality the interactions between the waves are not necessarily weak~\cite{zakharov1992kolmogorov, Kolmakov:2014}, and the increasing importance of quantum pressure in the GPE with increasing $k$. In the future, the experimental flexibility offered by atomic gases, in particular the possibility to tune the strength of nonlinearity via Feshbach resonances, might allow better understanding of the applicability of the approximate analytical predictions, and the limitations of the classical-field methods. 


\bibliographystyle{naturemag}

\begin{thebibliography}{99}
\bibitem{FootnoteSnapshots}
\bibinfo{note}{Note that snapshots of the turbulent flow do not obey the
  cylindrical symmetry of the (time-dependent) Hamiltonian. In real physical
  systems any such symmetry is always broken by imperfections; in our
  simulations the symmetry breaking is provided by the position of the
  numerical grid.}
\bibitem{KolmogorovSpectrumNote}
\bibinfo{note}{Note that this result is distinct from the prediction for the
  Kolmogorov energy spectrum of incompressible-flow dominated turbulence,
  $\mathcal{E}(k) \propto
  k^{-5/3}$~\cite{kolmogorov1941local,obukhov1941distribution}.}
\end{thebibliography}

\setlength{\parindent}{0pt}
\vspace{0.3cm}
\textbf{Acknowledgements} 
We thank G. V. Shlyapnikov, B. Svistunov, S. Stringari, N. R. Cooper, J. Dalibard, M. J. Davis, R. J. Fletcher, M. W. Zwierlein, K. Fujimoto and M. Tsubota for insightful discussions, and C. Eigen for experimental assistance. This work was supported by AFOSR, ARO, DARPA OLE, EPSRC [Grant No. EP/N011759/1], and ERC (QBox). The GeForce GTX TITAN X used for the numerical simulations was donated by the NVIDIA Corporation. N.N. and A.L.G. acknowledge support from Trinity College, Cambridge, and R.P.S. from the Royal Society. 

\textbf{Author Contributions} 
N.N. initiated the project, took and analysed the data. A.L.G. wrote the code for the numerical simulations and analysed the results. Z.H. supervised the project. All authors contributed extensively to the interpretation of the data, and the writing of the manuscript.

\textbf{Author Information} 
The authors declare no competing financial interests. Correspondence and requests for materials should be addressed to N.N. (nn270@cam.ac.uk).



\section{Methods}

\setcounter{section}{0}
\setcounter{subsection}{0}
\setcounter{figure}{0}
\renewcommand{\figurename}[1]{Extended Data Figure }
\makeatletter
\renewcommand{\thefigure}{\@arabic\c@figure} 
\makeatother
\renewcommand\thetable{S\arabic{table}}
\renewcommand{\vec}[1]{{\boldsymbol{#1}}}

{\bf Experimental system.} 
The BEC of $^{87}$Rb atoms is produced in a quasi-uniform potential of a can-shaped dark optical dipole trap (see Fig.~\ref{Fig1}). The repulsive trap walls are sculpted using 532~nm laser light and a phase-imprinting spatial light modulator. They are formed by one circular tube beam (propagating along $z$) and two thin sheet beams that act as end caps~\cite{gaunt2013bose}. 
At the end of evaporative cooling the trap depth is $\approx \kB\times 10$~nK and the condensed fraction is $\eta >0.9$.
We then slowly (over $700\,$ms) raise the trap depth to $U_0\approx \kB\times 60\,$nK, which does not result in any observable drop in $\eta$. 
Our atom number is calibrated to within 10\% from the critical temperature for condensation~\cite{schmidutz2014qjt}. 
The gradient of the modulus of the magnetic field along $z$, used to shake the cloud, is calibrated by pulsing it on for a short time $\delta t$ just after releasing the cloud from the trap and measuring the resulting velocity kick $\Delta U \delta t/(mL)$ in ToF. 

\vspace{2mm}
{\bf Phase-shift measurement of the resonance.}
The CoM position of the cloud in ToF is $v\, t_{\rm ToF}$, where $v$ is its velocity at the time of release.
In analogy with a driven damped harmonic oscillator, we assume that for a driving force $\propto \sin(\omega t)$ in steady state  $v(t) =A_\omega \omega \cos(\omega t+\phi_\omega)$, where $A_\omega$ and $\phi_\omega$ are  $\omega$-dependent (in-trap) displacement amplitude and phase shift. 
For a shaking time $t_S$ such that $\omega t_S$ is a multiple of $2\pi$, the response $v(t_S)=A_\omega \omega\cos(\phi_\omega)$ vanishes on resonance, and more generally
\begin{equation}\label{PhaseShiftEq}
v (t_S) \propto\frac{\omega(\omega^2-\omega_{\rm res}^2)}{(\omega^2-\omega_{\rm res}^2)^2+\Gamma^2\omega^2} \, ,
\end{equation}
where $\Gamma$ is the damping rate. 
In practice we fix $t_S=2\,$s and make measurements at discrete points $\omega = j\Delta\omega$, where $\Delta\omega=2\pi\times 0.5$~Hz and $j$ is an integer. We then use Eq.~(\ref{PhaseShiftEq}) to fit the data (see Figs.~\ref{Fig2}b and \ref{Fig2}d) with $\omega_{\rm res}$ and $\Gamma$ as free parameters.

\vspace{2mm}
{\bf Numerical methods.} 
We implement a three-dimensional numerical simulation of the Gross-Pitaevskii equation
\begin{equation}\label{FullGPE3D}
i\hbar \frac{\partial{\psi}}{\partial t}=\left(-\frac{\hbar^2}{2m}\nabla^2+V_\textrm{ext}({\bf r},t)+g|\psi|^2\right)\psi,
\end{equation}
where the coupling constant is $g=4\pi\hbar^2 a_s/m$, with $a_s$ the $s$-wave scattering length, and $V_\textrm{ext}({\bf r},t)$ is an external potential. 
We have $V_\textrm{ext}({\bf r},t)=V_\textrm{box}({\bf r})+V_\textrm{S}({\bf r},t)$, where $V_\textrm{box}({\bf r})$ is the (static) box-trap potential and  $V_\textrm{S}({\bf r},t)=\Delta U\sin(\omega t)z/L$ is the (time-varying) shaking potential. 
We perform numerical simulations on a cubic grid of $256^3$ points. Using a symmetrised split-step Fourier method we evolve the Bose field in time steps of $10\,\mu$s for up to $4\,$s. The calculations are performed at FP32 precision on an NVIDIA GeForce GTX TITAN X graphics card, and we achieve a running time of under 30~min for simulating each $1\,$s of dynamics.

\vspace{2mm}
{\bf Bogoliubov equations for the box trap.} 
The starting point for the analysis of the linear response of the BEC to external perturbations is the GPE in Eq.~(\ref{FullGPE3D}). We start by expanding $\psi({\bf r},t)$ around $\psi_0({\bf r})$, the ground state in the potential $V_\textrm{ext}({\bf r},t)=V_{\textrm{box}}({\bf r})$: 
\begin{equation}
\psi({\bf r},t)=e^{-i\mu t/\hbar} \left( \psi_0({\bf r})+u({\bf r})e^{-i\omega t}-v^*({\bf r})e^{i\omega t} \right) ,
\end{equation} 
where $^*$ denotes the complex conjugate. Linearising with respect to $u$ and $v$ leads to the Bogoliubov equations:
\begin{equation}
\left( \begin{array}{cc}
\mathcal{L} & -gn_0 \\
gn_0 & -\mathcal{L} \\
\end{array} \right)
\left( \begin{array}{c}
u \\
v\\
\end{array} \right)=\hbar\omega\left( \begin{array}{c}
u \\
v\\
\end{array} \right), \label{BogoliubovEqs}
\end{equation}
where $\mathcal{L}=-\frac{\hbar^2}{2m}\nabla^2+V_\textrm{box}({\bf r})-\mu+2gn_0({\bf r})$, and $n_0({\bf r})=|\psi_0({\bf r})|^2$ is the ground-state density. The set of eigenvalues $\hbar\omega$ forms the spectrum of elementary excitations, and $u$ and $v$ give the corresponding eigenmodes. For periodic boundary conditions $n_0$ is constant and one recovers the usual Bogoliubov spectrum~\cite{pitaevskii2003bose}. However, for fixed boundary conditions, $n_0({\bf r})$ is not analytical, and not even separable in the cylindrical coordinates. We solve the Bogoliubov equations by first computing $n_0({\bf r})$ from an imaginary-time propagation of the GPE, and then numerically solving Eq.~(\ref{BogoliubovEqs}). It is convenient to work in the basis of the free-particle eigenstates in a cylindrical box, so that the boundary conditions are automatically satisfied. Since we focus on the longitudinal modes, we restrict the basis to azimuthally-symmetric functions. Our results for the frequency of the lowest-lying (antisymmetric) mode along $z$ are shown in Fig.~\ref{Fig2}c as a grey shaded area (taking into account the uncertainty in the box size).
In the limit of vanishing shaking amplitude direct GPE simulations give the same results as the Bogoliubov approach.

\vspace{2mm}
{\bf Anisotropy analysis.} 
To quantify the anisotropy of the ToF expansion, seen in the atomic distribution after a long $t_{\rm ToF}$, we start with the column-density distribution in the ToF image, $\tilde{n}(y,z)=\int\textrm{d}x \; n({\bf r})$, where $n({\bf r})$ is the three-dimensional distribution and $x$ is the imaging axis. 
We then define the angular density distribution:
\begin{equation}\label{defptheta}
p(\theta)=\frac{1}{\bar{N}}\int_{r_1}^{r_2} r\textrm{d}r\;\tilde{n}(r\cos\theta,r\sin\theta) ,
\end{equation}
where the polar origin is set at the CoM of $\tilde{n}(y,z)$ and $\bar{N}$ is the total atom number in the shell $[r_1,r_2]$, so that $\int_0^{2\pi} \textrm{d}\theta \, p(\theta) = 1$. An isotropic distribution corresponds to the uniform $p_{\rm iso}(\theta)=(2\pi)^{-1}$. 
To quantify the anisotropy as a deviation from this uniform distribution, we introduce a simple heuristic measure:
\begin{equation}\label{DefAnisotropy}
\mathcal{A}=\frac{1}{2}\int_0^{2\pi}\textrm{d}\theta\;\left|p(\theta)-\frac{1}{2\pi} \right| ,
\end{equation}
so that $\mathcal{A}=0$ for $p_{\rm iso}$ and $\mathcal{A} \rightarrow 1$ for sharply-peaked distributions.
For our pure BEC, the diamond-shaped ToF distribution is close to the idealised square-shaped distribution depicted in the insets of Fig.~\ref{Fig1}b and \ref{Fig1}d, for which (taking $r_1=0$ and $r_2\rightarrow\infty$) the angular distribution is $p(\theta)= p_0(\theta \textrm{ mod } \pi/2)$, with
\begin{equation}
p_0(\theta)= \frac{1}{8} \left\{ \begin{array}{ll}
         \cos^{-2}\theta & \quad \mbox{for \quad $0\leq\theta <\pi/4$};\\
       \sin^{-2}\theta & \quad \mbox{for \quad $\pi/4 \leq \theta <\pi/2$}.\end{array} \right.  
\end{equation} 
For this distribution $\mathcal{A} \approx 9\%$. In experiments any imaging imperfections lead to a positive $\mathcal{A}$; for an equilibrium thermal gas ($\eta=0$), which is known to be isotropic, we observe a small residual $\mathcal{A} \approx 3\%$, and define all our experimental values of $\mathcal{A}$ from this baseline. 
For the inset of Fig.~\ref{Fig2}e, the radial integration in Eq.~(\ref{defptheta}) is performed in the shell defined by $r_1=\hbar k_{\rm low}  t_\textrm{ToF}/m$  and $r_2=\hbar k_{\rm high} t_\textrm{ToF}/m$ (corresponding to the vertical red lines in Fig.~\ref{Fig3}a).
 
\vspace{2mm}
{\bf Measurement of the momentum distribution.} The measurement of the momentum distribution using the time-of-flight (ToF) technique requires a kinetic-energy-dominated state. We assess the validity of this measurement by comparing it to Bragg spectroscopy~\cite{Stenger:1999}, which can provide a faithful measurement of the momentum distribution even if the interaction energy is dominant over the kinetic one. We shine onto the trapped cloud two off-resonant laser beams with wavevectors ${\bf k}_1$ and ${\bf k}_2$, detuned from each other by a frequency $\Delta\nu$, such that the resulting 1D Bragg-diffraction optical lattice is aligned with the axis of the box trap, ${\bf k}_1-{\bf k}_2\propto{\bf \hat{z}}$. The angle between the beams is such that the recoil energy of the diffracted atoms, $E_r\approx \kB\times 320$ nK, is larger than the trap depth ($U_0\approx \kB\times 60\,$nK), allowing the diffracted atoms to escape, and also much larger than the spread of energies in the trapped gas.  Measuring the fraction of diffracted atoms as a function of $\Delta\nu$ yields the 1D momentum distribution along ${\bf \hat{z}}$, $n_{1\textrm{D}}(k_z)$, which is simply related to the planar distribution $\tilde{n}$ (measured in ToF) by an Abel transform:
\begin{equation}
n_{1\textrm{D}}(k_z)=\int\textrm{d}k_y\: \tilde{n}\left(\sqrt{k_y^2+k_z^2}\right).  
\end{equation} 
We use a long ($20$~ms) and low-power Bragg pulse to minimise Fourier broadening while always keeping the diffracted fraction below $15\%$. To compare the ToF and Bragg measurements we integrate our ToF images along ${\bf \hat{y}}$.

\begin{figure} [h]
\includegraphics[width=\columnwidth]{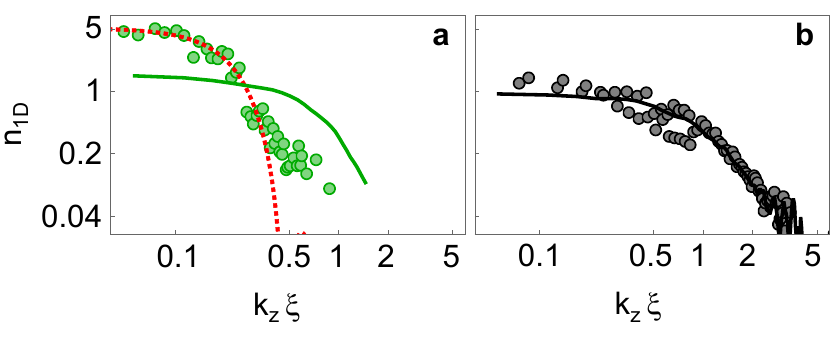}
\caption{
{\bf Momentum distributions: ToF versus Bragg.} Comparison of $n_{1\textrm{D}}(k_z)$ obtained using ToF expansion (solid lines) and Bragg spectroscopy (points), in the case of the initial quasi-pure BEC ({\bf a}) and the turbulent gas ({\bf b}). The red dashed line in {\bf a} corresponds to the Heisenberg-limited momentum distribution. All distributions are normalised to unity ($\int_0^{\infty}\textrm{d}(k_z\xi)n_{\textrm{1D}}=1$), without any adjustable parameter.
\label{FigBraggTOF}
}
\end{figure}

In Extended Data Fig.~\ref{FigBraggTOF}a, we apply both methods to the initial state, a quasi-pure BEC. In this case the ToF measurement (solid green line) overestimates the width of the momentum distribution, due to the importance of interactions during the expansion; the Bragg spectrum agrees well with the expected Heisenberg-limited distribution (dashed red line)~\cite{gotlibovych2014observing}. 
However, as shown in Extended Data Fig.~\ref{FigBraggTOF}b, in the relevant case of the kinetic-energy dominated turbulent state, the Bragg and ToF measurements are in excellent agreement, validating the assumptions made in the main text of the paper.  

\vspace{2mm}
{\bf Background-gas losses.}
In the absence of shaking, the atom population in the trap slowly decays due to collisions with the residual background gas in the vacuum chamber. These one-body losses are $k$-independent and described by an exponential decay with a time constant that we measured to be $\tau_{\textrm{vac}} =13$~s.
For analysing the population dynamics in the inset of Fig.~\ref{Fig3}b, we have corrected all populations for this background loss by multiplying them with a common factor $\exp(t_S/\tau_{\textrm{vac}})$.

\begin{figure} [h]
\includegraphics[width=\columnwidth]{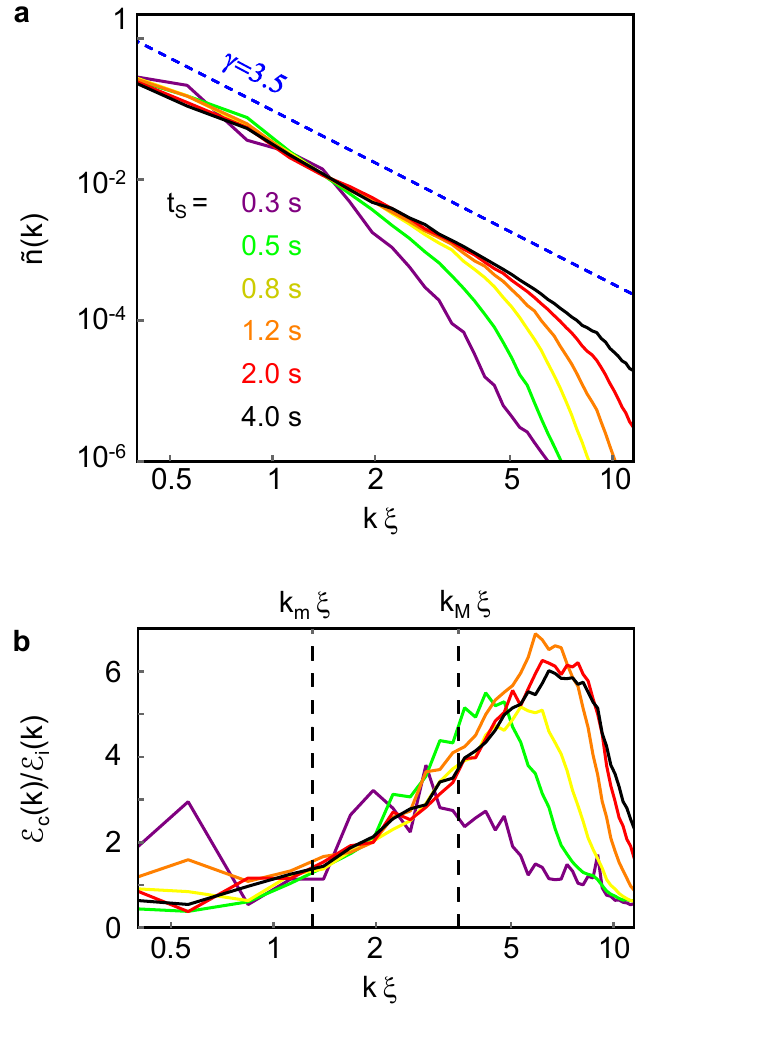}
\caption{
{\bf Turbulent cascade in numerical simulations.} {\bf a},  In-plane momentum distribution $\tilde{n}(k)$ for various shaking times. {\bf b}, Ratio of the compressible- ($\mathcal{E}_c$) to incompressible-flow ($\mathcal{E}_i$) components of the fluid-dynamical kinetic energy, with same colour code for shaking times as in {\bf a}. The simulation parameters for both panels are $N=8\times10^4$, shaking frequency $\omega/(2\pi)=9$~Hz, and shaking amplitude $\Delta U=\mu$.
\label{FigSimuTurb}
}
\end{figure}

\vspace{2mm}
{\bf Numerical simulations of the turbulent cascade.}  
In Extended Data Fig.~\ref{FigSimuTurb}a we show simulations of the dynamics of the in-plane momentum distribution in a shaken gas. Here $\tilde{n}(k)$ is computed from the spatial Fourier transform of the matter-wave field $\psi({\bf r},t_S)$. We observe that with increasing $t_S$ the same power-law behaviour gradually extends from large to ever smaller lengthscales, as expected for a direct cascade.

Following the procedure outlined in~\cite{nore1997kolmogorov}, we also study the fluid-dynamical kinetic-energy spectrum, $\mathcal{E}(k)$. 
We start by computing ${\tilde{\bf w}}({\bf k})$, the Fourier transform of the flow field
\begin{equation}
{\bf w}({\bf r})= (\hbar/m) \, |\psi({\bf r})| \, {\bf \nabla}\varphi({\bf r}) \, ,
\end{equation}
where $\varphi({\bf r})$ is the phase of $\psi({\bf r})$ and we omit the time label for brevity. 
Summing  $|{\tilde{\bf w}}({\bf k})|^2$ over all momenta with $|{\bf k}| = k$ gives the total $\mathcal{E}(k)$. Decomposing ${\tilde{\bf w}}({\bf k})$ into longitudinal and transverse components splits $\mathcal{E}(k)$ into, respectively, the compressible- ($\mathcal{E}_c$) and incompressible-flow ($\mathcal{E}_i$) contributions. In Extended Data Fig.~\ref{FigSimuTurb}b we plot the ratio $\mathcal{E}_c(k)/\mathcal{E}_i(k)$. We find that $\mathcal{E}_c(k)$ dominates in the $k$-range $k_m< k < k_M$, where the power-law momentum distribution is observed in both experiments and simulations; 
the same numerical observation was independently made by K. Fujimoto and M. Tsubota~\cite{osakaprivatecomm}. 
This supports the view that in the turbulent cascade waves play a more important role than vortices. Also note that vortices have core size $\sim \xi$, so it qualitatively makes sense that their contribution is not significant when $k\gtrsim 1/\xi$.


\end{document}